\documentclass[preprints,article,accept,moreauthors,pdftex]{my-mdpi}

\makeatletter							
\renewcommand{\maketag@@@}[1]{\hbox{\m@th\normalsize\normalfont#1}}%
\makeatother							
\pdfoutput=1

\firstpage{1} 
\pubvolume{11}
\issuenum{8}
\articlenumber{400}
\pubyear{2022}
\copyrightyear{2022}
\externaleditor{{Academic Editor: J. Alberto \linebreak Conejero}} 
\datereceived{15 July 2022} 
\daterevised{30 Jul, 3 and 9 Aug 2022} 
\dateaccepted{11 August 2022} 
\datepublished{13 August 2022} 
\hreflink{https://doi.org/\linebreak 10.3390/axioms11080400} 
\doinum{10.3390/axioms11080400}

\Title{Stability Analysis of Delayed COVID-19 Models}

\TitleCitation{Stability Analysis of Delayed COVID-19 Models}

\Author{Mohamed A. Zaitri $^{\dagger}$\orcidA{}, 
Cristiana J. Silva *$^{,\dagger}$\orcidB{} 
and Delfim F. M. Torres $^{\dagger}$\orcidC{}}

\AuthorNames{Mohamed A. Zaitri, Cristiana J. Silva and Delfim F. M. Torres}

\AuthorCitation{Zaitri, M.A.; Silva, C.J.; Torres, D.F.M.}

\address[1]{Center for Research and Development in Mathematics and Applications (CIDMA),
Department of Mathematics, University of Aveiro, 3810-193 Aveiro, Portugal}

\corres{\hangafter=1 \hangindent=1.05em \hspace{-0.82em}Correspondence: cjoaosilva@ua.pt}

\firstnote{\hangafter=1 \hangindent=1.05em \hspace{-0.82em}These authors contributed equally to this work.} 

\abstract{We analyze mathematical models for COVID-19 
with discrete time delays and vaccination.
Sufficient conditions for the local stability of the endemic 
and disease-free equilibrium points are proved 
for any positive time delay. The stability results 
are illustrated through numerical simulations performed in MATLAB.}



\begin{document}
	

\section{Introduction}
\label{sec:01}

A pandemic is an epidemic of an infectious disease occurring worldwide, 
or over a very wide area, crossing international boundaries and usually 
affecting a large number of people \cite{dic:pandemic}. An infectious 
disease, also known as transmissible disease or communicable disease, 
results from an infection caused by a large range of pathogens
such as, for example, bacteria and viruses. Several modes of 
transmission can be identified, such as droplet,  
fecal, sexual and/or oral  and vector-borne transmissions. 
Historically, communicable diseases have killed millions of people 
around the world, for example, smallpox, plague, great flu, polio, 
tuberculosis, HIV/AIDS, SARS, global H1N1 flu, cholera, measles, 
and ebola. Following the World Health Organization (WHO), some of 
these infectious diseases remain a threat for public health, 
such as cholera, measles, HIV, and tuberculosis 
\cite{who:outbreaks,who:hiv,who:tb}. 

In December 2019, a very dangerous SARS-CoV-2 virus quickly invaded 
the city of Wuhan in China and, subsequently, 183 countries in the world 
\cite{art1,art2}. WHO declared, on 30 January 2020, the COVID-19 infectious 
disease as a pandemic \cite{WHO}. It is now known that the spread of COVID-19 
changes very rapidly; therefore, taking appropriate and timely actions 
can influence the course of this pandemic. 

Since the beginning of the COVID-19 pandemic, researchers proposed 
different and complementary mathematical models that describe, 
approximately, the spread of SARS-CoV-2 in different 
regions of the world and with alternative modeling techniques; see, e.g., 
\cite{Cappi,Davis,LemosP,MR4200529,Zine,Tang:brazil,MyID:459}.
Although the literature dealing with models of COVID-19 is now huge,
with special issues  \cite{HERNANDEZVARGAS2021424},
books \cite{MyID:461}, and review papers \cite{MR4391308} on this topic,
deterministic models of COVID-19 with delay differential equations 
and vaccination are relatively scarce~\cite{MR4342396}.

The introduction of time delays to mathematical epidemic models 
has been studied in order to better understand and describe 
the transmission dynamics of infectious diseases; see, e.g., 
\cite{Arino,Zine,SilvaMaurerDelay,SilvaMaurerTorresDelay}. Moreover, 
time delays may have an important effect on the stability of the 
equilibrium points, leading, for example, to periodic solutions 
by Hopf bifurcation; see, e.g., \cite{Tipsri:Chaos:2015}  
and references cited therein.  

As in other infectious diseases, the latent and incubation periods 
have an important role in the spread of COVID-19. The latent period 
of an infectious disease is the time interval between infection and 
becoming infectious, whether the incubation period is the time interval 
between infection and the appearance of clinical symptoms 
\cite{Fine:AmJEpid,He:temporal:covid,Xin:latent:covid}. 
Following the WHO, the incubation period for COVID-19 is between 
2 and 10 days \cite{art5}. In~\cite{Xin:latent:covid}, the authors 
estimated the mean latent period to be 5.5 (95\% CI: 5.1–5.9) days, 
shorter than the mean incubation period (6.9 days).  
However, and differently from other infectious diseases, 
asymptomatic infected individuals can transmit the infection 
and this imposes more strict mitigation strategies; see, e.g., 
\cite{Muller:Lancet:2021}. To describe and analyze this biological 
phenomenon, we generalize here a compartmental mathematical model, 
first proposed in~\cite{Pengetal}, by considering a system of delayed 
differential equations with discrete time delays. 

In  recent years, several epidemic models have been presented, both
stochastic and deterministic ones; see e.g., \cite{Calleri,Rihan}.
In \cite{Pengetal}, a deterministic mathematical model is proposed 
to analyze the spread of the COVID-19 epidemic, based on a dynamic 
mechanism that incorporates the intrinsic impact of hidden latent 
and infectious cases on the entire process of virus transmission. 
In \cite{zaitri}, Zaitri et al. applied optimal control theory to a 
generalized SEIR-type model, based on \cite{Pengetal}, with three controls, 
representing social distancing, preventive means, and treatment measures 
to combat the spread of the COVID-19 pandemic. They analyzed such optimal 
control problem with respect to real data transmission in Italy.  
Their results show the appropriateness of the model, in particular 
with respect to the number of quarantined/hospitalized (confirmed 
and infected) and recovered individuals.
Alternative approaches based on SIR-type models but 
that combine machine learning methods have also been developed;
see, e.g., \cite{Lozano,Miikkulainen}.

In our paper, we modify the model analyzed in \cite{Pengetal} 
in order to consider time delays, birth and death rates. More precisely, 
we introduce a time delay that represents, mathematically, the fact that 
the migration of individuals from susceptible to infected is subject 
to  delay. Secondly, we present a normalized version of the SEIR-type model, 
compute the equilibrium points, the basic reproduction number, and we prove 
sufficient conditions for the stability of the equilibrium points, 
for any positive time delay. Then, we extend the previous model in order 
to consider vaccination and perform numerical simulations taking into account the 
real data of the spread of COVID-19 in Italy from 18 October 2020, 
to 17~January 2021. This allows us to compare our results with previous ones.

The paper is organized as follows:~In Section~\ref{sec:model1}, we propose 
a delayed $SEIQRP$ mathematical model for COVID-19. Considering the normalized 
model of the delayed $SEIQRP$ model, we prove sufficient conditions 
for the stability of the equilibrium points for any time delay. Then, in 
Section~\ref{section:vaccine:constant}, we propose a delayed mathematical 
model for COVID-19 with vaccination. Analogously, we prove sufficient conditions 
for the stability of the equilibrium points of the normalized $seiqrpw$
with vaccination, for any time delay. Numerical simulations and a discussion
of the results are provided in Section~\ref{sec:numerical}, illustrating the stability 
of both delayed models and their practical utility.


\section{The Delayed SEIQRP Model}
\label{sec:model1}

In this section, we propose a delayed mathematical model for COVID-19, 
which generalizes the one proposed in \cite{Pengetal}. 
As mentioned in Section~\ref{sec:01}, there are many 
different models but, all of them, are approximations of the reality. 
For example, in \cite{Giordano}
the possibility to become susceptible again is ignored,
although we know re-infection is possible and occurs; while in
\cite{Liu} deaths are not taken into account.

Our model considers six state variables: 
susceptible individuals, $S(t)$; exposed individuals, $E(t)$; 
infected individuals, $I(t)$; quarantined individuals, $Q(t)$; 
recovered individuals, $R(t)$; and insusceptible/protected individuals, $P(t)$. 
The total population is denoted by $N(t)$ and is given by 
\begin{equation}
\label{totalpolpulation:N}
N(t)= S(t)+E(t)+I(t)+Q(t)+R(t)+P(t), 
\quad \text{ for all } \quad  t \in [0, T]. 
\end{equation}

The following assumptions are made to describe the spread of COVID-19: 
$b$ is the birth rate, $\mu$ is the death rate, $\alpha$ is the protection rate, 
$\beta$ the infection rate, $\gamma$ the inverse of the average latent time, 
$\delta$ the rate at which infectious people enter in quarantine, and
$\lambda$ the recovery rate. The time delay $\tau \geq 0$ represents 
the incubation period, that is, the length of time before 
the infected individuals become infectious.

We introduce a discrete time delay that represents the transfer delay 
from the class of susceptible individuals to the class of infected individuals, 
after the contact of a susceptible individual with an infectious one. 
Precisely, the model we propose is given by the following system 
of six nonlinear ordinary delayed differential equations:
\begin{equation}
\label{model}
\begin{cases}
\dot{S}(t) = b N(t) - \frac{\beta S(t-\tau) I(t-\tau)}{N(t)} 
- (\alpha +\mu) S(t),\\[0.2 cm]
\dot{E}(t) = \frac{\beta S(t-\tau) I(t-\tau)}{N(t)} 
- (\gamma + \mu)E(t),\\[0.2 cm]
\dot{I}(t) = \gamma E(t) - (\delta + \mu) I(t),\\[0.2 cm]
\dot{Q}(t) = \delta I(t) - (\lambda + \mu) Q(t),\\[0.2 cm]
\dot{R}(t) = \lambda Q(t) - \mu R(t),\\[0.2 cm]
\dot{P}(t) = \alpha S(t) - \mu P(t), 
\end{cases}
\end{equation}
where the state variables are subject to the initial conditions
$S(\theta) = S_{0}$, $\theta\in [-\tau,0]$, $E(0)=E_0$, $I(\theta) = I_{0}$,
$\theta\in [-\tau,0]$, $Q(0) = Q_{0}$, $R(0)=R_0$, and $P(0) = P_{0}$.


\subsection{The Normalized $seiqrp$ Delayed Model}
\label{subsec:normalize:delay}

In the situation where the total population size $N(t)$ is not constant along time, 
it is often convenient to consider the proportions of each compartment 
of individuals in the population, namely
$s(t)=\frac{S(t)}{N(t)}$, $e(t)=\frac{E(t)}{N(t)}$, $i(t)=\frac{I(t)}{N(t)}$, 
$q(t)=\frac{Q(t)}{N(t)}$, $r(t)=\frac{R(t)}{N(t)}$, and $p(t)=\frac{P(t)}{N(t)}$.
According to equality \eqref{totalpolpulation:N}, we have $\dot{N}(t)=(b-\mu)N(t)$. 
Therefore, the normalized $seiqrp$ delayed model is given by
\begin{equation}
\label{model:seir:normalized}
\begin{cases}
\dot{s}(t) = b- \beta\, s(t-\tau)\,i(t-\tau)- (\alpha + b)\, s(t)\, , \\[0.2 cm]
\dot{e}(t) = \beta\, s(t-\tau)\,i(t-\tau) - (\gamma + b)\,e(t)\, , \\[0.2 cm]
\dot{i}(t) = \gamma\, e(t) - (\delta + b)\, i(t) \, , \\[0.2 cm]
\dot{q}(t) = \delta\, i(t) - (\lambda + b)\, q(t)\, , \\[0.2 cm]
\dot{r}(t) = \lambda\, q(t) - b\, r(t) \, , \\[0.2 cm]
\dot{p}(t) = \alpha \,s(t) - b\, p(t) \, .
\end{cases}
\end{equation}
The state variables for system \eqref{model:seir:normalized} are subject 
to the following initial conditions: $s(\theta) = \dfrac{S_{0}}{N(0)}$,
$\theta\in [-\tau,0]$, $e(0)=\dfrac{E_0}{N(0)}$, $i(\theta) = \dfrac{I_{0}}{N(0)}$,
$\theta\in [-\tau,0]$, $q(0) = \dfrac{Q_{0}}{N(0)}$, $r(0)=\frac{R_0}{N(0)}$, 
and $p(0) = \dfrac{P_{0}}{N(0)}$, with $s(t) + e(t) + i(t) + q(t) + r(t) + p(t) = 1$. 

In Section~\ref{sec:2.2} we show that model \eqref{model:seir:normalized} 
has two equilibrium points: the disease free and the endemic equilibrium. 


\subsection{Equilibrium Points and the Basic Reproduction Number}
\label{sec:2.2}

The disease free equilibrium and the endemic equilibrium point 
are obtained by solving the right hand side of equations 
in \eqref{model:seir:normalized} equal to zero:\vspace{6pt}
\begin{align*}
b- \beta\, s(t-\tau)\,i(t-\tau)- (\alpha + b)\, s(t) &=0\, , \\
\beta\, s(t-\tau)\,i(t-\tau) - (\gamma + b)\, e(t)	&=0\, , \\
\gamma\, e(t) - (\delta + b)\, i(t)	&=0\, , \\
\delta\, i(t) - (\lambda + b)\, q(t) &=0\, ,\\
\lambda\, q(t) - b\, r(t) &=0\, , \\
\alpha\, s(t) - b\, p(t) &=0\, ,
\end{align*}
from which the disease free equilibrium, $\Sigma_0$, is given by
\begin{equation}
\label{eq:DFE}
\Sigma_0 = \left(s_{0}, e_0, i_{0}, q_{0}, r_{0}, p_0  \right) 
= \left(\frac{b}{\alpha+b} , 0, 0, 0, 0, \frac{\alpha}{ \alpha+b}\right),
\end{equation}
while the endemic equilibrium point, $\Sigma^+$, is given by
\begin{equation}
\label{eq:EE}
\Sigma^+ = \left(s^+, e^+, i^+, q^+, r^+, p^+  \right)
\end{equation}
with
\begin{equation}
\begin{split}
s^+ &= \frac { \left(\delta+b \right)\,\left(\gamma+b \right)}{\beta\, \gamma},\\
e^+ &= \frac{\beta\,s^+\,i^+}{\left( \gamma+b \right)},\\
i^+ &= \frac{\beta\,\gamma\,b-\left( \delta+b \right)  
\left( \gamma+b \right)  \left( \alpha+b \right)}{\beta\,
\left( \delta+b \right)\,  \left( \gamma+b \right)},\\
q^+ &=\frac{\beta\,\gamma\,b\, \delta-\delta\,
\left( \delta+b \right)  \left( \gamma+b \right)  
\left( \alpha+b \right)}{\beta\,\left( \lambda+b \right)\,
\left( \delta+b \right)\,  \left( \gamma+b \right)},\\
r^+ &= \frac{\lambda\,\delta\,\beta\,\gamma\,b
-\lambda\,\delta\,\left( \delta+b \right)\left( \gamma+b \right)  
\left( \alpha+b \right)}{b\,\beta\,\left( \lambda+b \right)
\,\left( \delta+b \right)\,  \left( \gamma+b \right)},\\
p^+ &= \frac { \alpha\,\left(\delta+b \right)
\,\left(\gamma+b \right)}{b\,\beta\, \gamma}.
\end{split}
\end{equation}

Following the method of van den Driessche \cite{MR1950747}, 
one easily compute the following basic reproduction number: 
\begin{equation}
\label{eq:R0}
R_0 = \frac {\beta\,\gamma\,b}{ \left( \alpha
+b \right) \left( \delta+b \right)  
\left( \gamma+ b \right)  } \, .
\end{equation}

The reader interested in the details of the algorithm
according to which the basic reproduction number \eqref{eq:R0}
is computed, is referred to the open access article \cite{MyID:417}.


\subsection{Stability of the Normalized $seiqrp$ Delayed Model}
\label{stability:01}

Now, we prove some sufficient conditions for the local asymptotic 
stability of the disease free equilibrium, $\Sigma_0$, and the endemic 
equilibrium point, $\Sigma^+$, for any time delay $\tau \geq 0$. 

Consider the following coordinate transformation:
$x_1(t)=s(t)-\bar{s}$, $x_2(t)=e(t)-\bar{e}$, 
$x_3(t)=i(t)-\bar{i}$, $x_4(t)=q(t)-\bar{q}$, 
$x_5(t)=r(t)-\bar{r}$, and $x_6(t)=p(t)-\bar{p}$,
where $(\bar{s},\bar{r},\bar{i},\bar{q},\bar{r},\bar{p})$ 
denotes any equilibrium point of system \eqref{model:seir:normalized}. 
The linearized system of \eqref{model:seir:normalized} takes the form
\begin{equation}
\dot{X}(t)=A_0\,X(t)+A_1\,X(t-\tau),
\end{equation} 
where $X=(x_1,x_2,x_3,x_4,x_5,x_6)^T$,\vspace{6pt}
\begin{equation*}
A_0=\begin{pmatrix}
-\alpha-b & 0 & 0 & 0 & 0 & 0\\
0 & -\gamma-b & 0 & 0 & 0 & 0 \\
0 & \gamma & -\delta-b & 0 & 0 & 0\\ 
0 & 0 & \delta & -\lambda-b & 0 & 0\\
0 & 0 & 0 & \lambda & -b &0  \\
\alpha & 0 & 0 & 0 & 0 &-b
\end{pmatrix},
\end{equation*}
and	
\begin{equation*}
A_1=\begin{pmatrix}
-\beta\,\bar{i} & 0 & -\beta\,\bar{s} & 0 & 0 & 0\\
\beta\,\bar{i} & 0 &  \beta\,\bar{s} & 0 & 0 &0\\
0 & 0 & 0 & 0 & 0 & 0 \\
0 & 0 & 0 & 0 & 0 & 0 \\
0 & 0 & 0 & 0 & 0 & 0 \\
0 & 0 & 0 & 0 & 0 & 0 
\end{pmatrix}. 
\end{equation*}

The characteristic equation of system \eqref{model:seir:normalized}, 
for any equilibrium point, is given by
\begin{equation}
\label{caract}
\Delta(y)=\vert y\, Id_{6\times 6}-A_0-A_1\,e^{-\tau\,y}\vert.
\end{equation}

We are now in a position to prove our first two results.

\begin{Theorem}[Stability of the disease free equilibrium of system \eqref{model:seir:normalized}]
If $R_0 < 1$, then the disease free equilibrium $\Sigma_0$ 
is locally asymptotically stable for any time-delay $\tau\geq 0$. 
If $R_0>1$, then the disease free equilibrium $\Sigma_0$ 
is unstable for any time-delay $\tau\geq 0$.
\end{Theorem}

\begin{proof}
The characteristic equation of \eqref{model:seir:normalized}, 
at the disease free equilibrium $\Sigma_0$, is given by
\begin{equation}
\label{characteristic:equation}
P(y,\tau)=(y+b)^2\,(y+\alpha+b)\,(y+\lambda+b)
\,(y^2+\Lambda_1\,y+\Lambda_2(y))=0,
\end{equation}
where
$\Lambda_1=\delta+2\,b+\gamma$ and 
$\Lambda_2(y)=(\delta+b)\,(\gamma+b)
-\dfrac{\beta\,\gamma\,b}{\alpha+b }e^{-\tau \,y}$.	

Let $R_0 < 1$. We divide the proof into the non-delayed and delayed cases.	
\begin{itemize}
\item[(i)] Let $\tau=0$. In this case, 
the Equation \eqref{characteristic:equation} becomes
\begin{small}
\begin{equation}
\label{characteristic:equation:tau0}
P(y,0)=(y+b)^2\,(y+\alpha+b)\,(y+\lambda+b)
\,\left( y^2+\Lambda_1\,y+(\delta+b)
\,(\gamma+b)-\dfrac{\beta\,\gamma\,b}{\alpha+b }\right) =0 \, .
\end{equation}
\end{small}
\vspace{-3pt}

\noindent We need to prove that all roots of the characteristic Equation 
\eqref{characteristic:equation:tau0} have negative real parts. 
It is easy to see that $y_1=-b$, $y_2=-\alpha-b$ and $y_3=-\lambda-b$ 
are roots of Equation~\eqref{characteristic:equation:tau0} and all 
of them are real negative roots. Thus, we just need to analyze the fourth 
term of \eqref{characteristic:equation:tau0}, here denoted by $P_1$, that is,  
\begin{equation*}
P_1(y,0):=y^2+\Lambda_1\,y+(\delta+b)\,(\gamma+b)
-\dfrac{\beta\,\gamma\,b}{\alpha+b}. 	
\end{equation*}
Using the Routh--Hurwitz criterion \cite{Rogers:Routh:Hurwitz}, 
we know that all roots of $P_1(y,0)$ have negative real parts if, 
and only if, the coefficients of $P_1(y,0)$ are strictly positive. 
In this case, we have $\Lambda_1=\delta+2\,b+\gamma >0$ and  
$$
(\delta+b)\,(\gamma+b)-\dfrac{\beta\,\gamma\,b}{\alpha+b } >0 
\quad \textrm{ if and only if}\quad 
R_0= \frac {\beta\,\gamma\,b}{ \left( \delta+b \right)  
\left( \gamma+ b \right)  \left( \alpha+b \right) }<1.
$$
Therefore, we have just proved that the disease free equilibrium, $\Sigma_0$, 
is locally asymptotically stable for $\tau=0$, whenever $R_0 < 1$.
		
\item[(ii)] Let $\tau>0$. In this case, we will use 
Rouch\'e's theorem \cite{MR1218880,MR1880658}
to prove that all roots of the characteristic Equation 
\eqref{characteristic:equation} cannot intersect the imaginary axis, 
i.e., the characteristic equation cannot have pure imaginary roots.
Suppose the contrary, that is, suppose there exists $w\in \mathbb{R}$ 
such that $y = w\,i$ is a solution of \eqref{characteristic:equation}.
Replacing $y$ in the fourth term of \eqref{characteristic:equation}, 
we get that
\begin{equation*}
-w^2+(\delta+2\,b+\gamma)\,w\,i+(\delta+b)\,(\gamma+b)
-\dfrac{\beta\,\gamma\, b}{\alpha+b } 
\left(\cos (\tau\,w)-i\,\sin (\tau\,w)\right)=0.
\end{equation*}
Then,
\begin{equation*}
\begin{cases}
-w^2+(\delta+b)\,(\gamma+b) = \dfrac{\beta\, \gamma\,b}{\alpha+b}\,
\cos(\tau\,w),\\[0.2 cm]
(\delta+2\,b+\gamma)\,w =-\dfrac{\beta\,\gamma\,b}{\alpha+b}
\,\sin (\tau\,w) \, .
\end{cases}
\end{equation*}
By adding up the squares of both equations, 
and using the fundamental trigonometric formula, 
we obtain that
\begin{equation*}
w^4+\left( \left( \delta+b\right) ^2
+\left( \gamma+b\right) ^2\right) w^2+(\delta+b)^2\,(\gamma+b)^2
-\left(\dfrac{\beta\,\gamma\,b}{\alpha +b}\right)^2=0,
\end{equation*} 
which is equivalent to
\begin{equation}
w^2= \dfrac{1}{2} \sqrt{\left( \left( \delta+b\right)^2
-\left( \gamma+b\right) ^2\right)^2  
+4 \,\left(\dfrac{\beta\,\gamma\,b}{\alpha +b}\right)^2}
-\dfrac{1}{2}\left( \left( \delta+b\right)^2
+\left( \gamma+b\right) ^2\right).
\end{equation}
If $R_0<1$, then 
$(\delta+b)^2\,(\gamma+b)^2
-\left(\dfrac{\beta\,\gamma\,b}{\alpha +b}\right)^2 >0$, and 	
\begin{small}
$$
\left( \left( \delta+b\right) ^2+\left( \gamma+b\right) ^2\right)^2
-4\left( (\delta+b)^2\,(\gamma+b)^2
-\left(\dfrac{\beta\,\gamma\,b}{\alpha +b}\right)^2 \right) 
<\left( \left( \delta+b\right) ^2+\left( \gamma+b\right) ^2\right)^2,
$$
\end{small}
so that
$$
\sqrt{\left( \left( \delta+b\right) ^2-\left( \gamma+b\right) ^2\right)^2  
+4 \,\left(\dfrac{\beta\,\gamma\,b}{\alpha +b}\right)^2}
<\left( \delta+b\right) ^2+\left( \gamma+b\right) ^2.
$$
Hence, we have $w^2<0$, which is a contradiction. 
Therefore, we have proved that whenever
$R_0 < 1$, the characteristic Equation \eqref{characteristic:equation} 
cannot have pure imaginary roots and the disease free equilibrium $\Sigma_0$ 
is locally asymptotically stable, for any strictly positive time-delay $\tau$.

\item[(iii)] Suppose now that $R_0 > 1$. We know that the characteristic Equation 
\eqref{characteristic:equation} has three real negative roots $y_1=-b$, 
$y_2 =-\alpha-b$, and $y_3=-\lambda-b$. Thus, we need to check if the 
remaining roots of
\begin{equation}
q(y):=y^2+\Lambda_1\,y+\Lambda_2(y)
\end{equation}
have negative real parts. It is easy to see that $q(0)=\Lambda_2(0)<0$
because we are assuming $R_0 > 1$. On the other hand,
$\lim\limits_{y\rightarrow +\infty} q(y)=+\infty$. Therefore, 
by continuity of $q(y)$, there is at least one positive root 
of the characteristic Equation \eqref{characteristic:equation}. 
Hence, we conclude that $\Sigma_0$ is unstable when $R_0 > 1$.
\end{itemize}
The proof is complete.
\end{proof}

\begin{Theorem}[Stability of the endemic equilibrium point of system \eqref{model:seir:normalized}] 
\label{theo2}
Let $\tau= 0$. If $R_0>1$, then the endemic equilibrium point 
$\Sigma^+$ is locally asymptotically stable. When $\tau > 0$, 
the endemic equilibrium point $\Sigma^+$ is locally asymptotically stable 
if the basic reproduction number $R_0$ satisfies the following relations:
\begin{equation}
\label{R0:condition:1}
1<R_0<\min{ \left( 3,1+\dfrac{\sqrt{(\alpha+b)^2
+(\delta+b)^2+(\gamma+b)^2}}{\alpha+b} \right)} 
\end{equation}
and 
\begin{equation}
\label{R0:condition:2}
M_1 R_0^2+M_2 R_0+M_3>0,
\end{equation}
where
\begin{equation*}
\begin{split}
M_1&=-(\alpha+b)^2\,\left((\delta+b)^2+(\gamma+b)^2
\right),\\ 
M_2&=2\,\left( \alpha+b\right)^2\left(\left( \delta+\gamma
+2\, b\right) ^2-3\,(\delta+b)\,(\gamma+b)\right)\\  
&\qquad +2\,\left( \alpha+b\right) \,(\delta+b)\,(\gamma+b)\,(\delta+\gamma+2\, b),\\
M_3&=2\,(\alpha+b)\,(\delta+b)\,(\gamma+b)\left( \alpha-\delta-\gamma-b  \right).
\end{split}
\end{equation*}
\end{Theorem}

\begin{proof}
The characteristic Equation \eqref{caract}, computed 
at the endemic equilibrium point $\Sigma^+$, is given by
\begin{equation}
\label{characteristic:equation:EE}
\tilde{P}(y,\tau)=(y+b)^2\,(y+\lambda+b)\,
(y^3 +\Delta_1(y)\, y^2+\Delta_2(y)\,y+\Delta_3(y))=0 \, ,
\end{equation}
where
$\Delta_1(y)=L_1+\bar{L}_1\,e^{-\tau\,y}$,
$\Delta_2(y)=L_2+\bar{L}_2\,e^{-\tau\,y}$, 
and $\Delta_3(y)=L_3+\bar{L}_3\,e^{-\tau\,y}$
with 
\begin{equation*}
\begin{split}
L_1&=\alpha+\delta+\gamma+3\,b,\\ 
\bar{L}_1&=	\frac{\beta\,\gamma\,b-\left( \delta+b \right)  
\left( \gamma+b \right)  \left( \alpha+b \right)}{\left( 
\delta+b \right)\, \left(\gamma+b\right)},\\ 
L_2&=(\gamma+2\,b+\delta)\,(\alpha+b)+(\gamma+b)\,(\delta+b),\\ 
\bar{L}_2&= (\gamma+2\,b+\delta)\,(\alpha+b)\,(R_0-1)-(\gamma+b)\,(\delta+b),\\
L_3&=(\alpha+b)\,(\gamma+b)\,(\delta+b),\\  
\bar{L}_3&=\beta\,\gamma\,b-2\,(\alpha+b)\,(\gamma+b)\,(\delta+b).
\end{split}
\end{equation*}	
\begin{itemize}
\item[(i)] Let $\tau=0$. In this case, the Equation 
\eqref{characteristic:equation:EE} becomes
\begin{equation}
\label{characteristic:equation:tau0:EE}
\tilde{P}(y,0)=(y+b)^2\,(y+\lambda+b)
\,(y^3 +\tilde{\Delta}_1\, y^2+\tilde{\Delta}_2\,y+\tilde{\Delta}_3)=0,
\end{equation}
where
$\tilde{\Delta}_1=L_1+\bar{L}_1$, $\tilde{\Delta}_2=L_2+\bar{L}_2$ 
and $\tilde{\Delta}_3=L_3+\bar{L}_3$. We need to prove that all the 
roots of the characteristic Equation \eqref{characteristic:equation:tau0:EE} 
have negative real parts. It is easy to see that $y_1=-b$ and $y_2=-\lambda-b$ 
are roots of \eqref{characteristic:equation:tau0:EE} and both 
are real negative roots. Thus, we just need to consider the third term of the
above equation. Let
\begin{equation}
\label{P:tilde:star}
\tilde{P}_3(y,0):= y^3 +\tilde{\Delta}_1\, y^2
+\tilde{\Delta}_2\,y+\tilde{\Delta}_3=0 \, .	
\end{equation}
Using the Routh--Hurwitz criterion \cite{Rogers:Routh:Hurwitz}, 
we know that all roots of $\tilde{P}_3(y,0)$ have negative real parts 
if, and only if, the coefficients of $\tilde{P}_3(y,0)$ are strictly 
positive and 
$\tilde{\Delta}^*=\tilde{\Delta}_1\,\tilde{\Delta}_2-\tilde{\Delta}_3\,>0$. 
If $R_0>1$, then
\begin{equation*}
\begin{split}
\tilde{\Delta}_1&=\alpha+\delta+\gamma+3\,b+(\alpha+b)\,
(R_0-1) >0 ,\\
\tilde{\Delta}_2&=(\delta+\gamma+2\,b)\,(\alpha+b)\, R_0 >0,	\\
\tilde{\Delta}_3&=(\alpha+b)\,(\delta+b)\,(\gamma+b)
\,(R_0-1) >0,\\
\tilde{\Delta}^*&=(\alpha+b)\, (\alpha+b)\,(\delta+\gamma+2\,b)\,R_0^2\\
&\quad +(\alpha+b)\,(\delta^2+3\,b\,(\delta+b)+\gamma\,(\delta+\gamma+3\,b))\,R_0\\
&\quad +(\alpha+b)\,(\delta+b)\,(\gamma+b) > 0. 
\end{split}
\end{equation*}
		
\item[(ii)] Let $\tau>0$. Using Rouch\'e's theorem, we prove that all the roots 
of the characteristic Equation \eqref{characteristic:equation:EE} cannot 
intersect the imaginary axis, i.e., the characteristic equation cannot 
have pure imaginary roots. Suppose the opposite, that is, assume there exists 
$w\in \mathbb{R}$ such that $y = w\,i$ is a solution of \eqref{characteristic:equation:EE}.
Replacing $y$ into the third term of \eqref{characteristic:equation:EE}, we get that\vspace{6pt}
\begin{equation*}
-w^3\,i-L_1\,w^2+L_2\,w\,i+L_3 
+ (-\bar{L}_1\,w^2+\bar{L}_2\,w\,i+\bar{L}_3)\,
\left(\cos (\tau\,w)-i\,\sin (\tau\,w)\right)=0.
\end{equation*}
Then,
\begin{equation*}
\begin{cases}
-L_1\,w^2+L_3=(\bar{L}_1\,w^2-\bar{L}_3)\,\cos(\tau\,w)
-\bar{L}_2\,w\,\sin (\tau\,w),\\[0.2 cm]
-w^3+L_2\,w =- \bar{L}_2\,w \,\cos(\tau\,w)
- (\bar{L}_1\,w^2-\bar{L}_3)\,\sin (\tau\,w) \, .
\end{cases}
\end{equation*}
By adding up the squares of both equations, and using 
the fundamental trigonometric formula, we obtain that 
\begin{equation*}
w^6+K_1\,w^4+K_2\,w^2+K_3 =0,
\end{equation*}
where
\begin{equation*}
\begin{split}
K_1&=L_1^2-\bar{L}_1^2-2\,L_2 ,\\
K_2&=2\,\bar{L}_1\,\bar{L}_3-2\,L_1\,L_3
+L_2^2-\bar{L}_2^2,\\ 
K_3&=L_3^2-\bar{L}_3^2. 
\end{split}
\end{equation*}

Assume that the basic reproduction number $R_0$ satisfies relations 
\eqref{R0:condition:1} and \eqref{R0:condition:2} 
with the following condition:
\begin{small}
\begin{equation}
\label{condition:2}
\min{ \left( 3,1+\dfrac{\sqrt{(\alpha+b)^2+(\delta+b)^2
+(\gamma+b)^2}}{\alpha+b} \right)}
=1+\dfrac{\sqrt{(\alpha+b)^2+(\delta+b)^2
+(\gamma+b)^2}}{\alpha+b}.
\end{equation}	
\end{small} 
Then, 
$$
K_1=(\delta+b)^2+(\gamma+b)^2+(\alpha+b)^2\,
\left(1-\left(R_0-1\right)^2 \right)>0.
$$
In contrast, if $R_0$ satisfies 
relations \eqref{R0:condition:1} and \eqref{R0:condition:2}
with the condition
\begin{equation}
\label{condition:1}
\min{ \left( 3,1+\dfrac{\sqrt{(\alpha+b)^2
+(\delta+b)^2+(\gamma+b)^2}}{\alpha+b} \right)}=3 \, ,
\end{equation}
then we have	
\begin{equation*}
1<R_0<3< 1+\dfrac{\sqrt{(\alpha+b)^2+(\delta+b)^2+(\gamma+b)^2}}{\alpha+b},
\end{equation*}
which is equivalent to
\begin{equation*}
\begin{split}
&0<R_0-1<2<\dfrac{\sqrt{(\alpha+b)^2+(\delta+b)^2+(\gamma+b)^2}}{\alpha+b},\\
&1-\left(\dfrac{(\alpha+b)^2+(\delta+b)^2+(\gamma+b)^2}{(\alpha+b)^2}\right) <1-(R_0-1)^2<1,\\
&-(\delta+b)^2-(\gamma+b)^2<(\alpha+b)^2\left( 1-(R_0-1)^2\right) <(\alpha+b)^2.
\end{split}
\end{equation*}
Thus,
\begin{equation*}
K_1>0.
\end{equation*}
Under the assumption that the basic reproduction number $R_0$ satisfies 
relations \eqref{R0:condition:1} and \eqref{R0:condition:2}, we have
\begin{equation*}
K_2=M_1 R_0^2+M_2 R_0+M_3>0 \, .
\end{equation*}

Therefore, if we assume that the basic reproduction number $R_0$ satisfies 
relations~\eqref{R0:condition:1} and \eqref{R0:condition:2}
with condition \eqref{condition:1}, then
\begin{equation*}
K_3=(\alpha+b)^2\,(\delta+b)^2\,(\gamma+b)^2\,\left(1-\left(R_0-2\right)^2 \right)>0;
\end{equation*}
if $R_0$ satisfies relations \eqref{R0:condition:1} and \eqref{R0:condition:2}
with condition \eqref{condition:2}, then we have
\begin{equation*}
1<R_0<	1+\dfrac{\sqrt{(\alpha+b)^2+(\delta+b)^2+(\gamma+b)^2}}{\alpha+b}<3,
\end{equation*}
which is equivalent to
\begin{equation*}
-1<R_0-2<	-1+\dfrac{\sqrt{(\alpha+b)^2+(\delta+b)^2+(\gamma+b)^2}}{\alpha+b}<1,
\end{equation*}
and also equivalent to
\begin{equation*}
1-\left( R_0-2\right)^2>0.
\end{equation*}
Thus,
\begin{equation*}
K_3>0 \, .
\end{equation*}

We conclude that the left hand-side of equation \eqref{characteristic:equation:EE} 
is strictly positive, which implies that this equation is not possible. Therefore,
\eqref{characteristic:equation:tau0:EE} does not have imaginary roots,	
which implies that $\Sigma^+$ is locally asymptotically 
stable for any time delay $\tau > 0$.
\end{itemize}
The proof is complete.
\end{proof}	

It should be noted that Theorem~\ref{theo2} is not trivial, 
and it is not easy to give a biological/medical interpretation 
to the relations \eqref{R0:condition:1} and \eqref{R0:condition:2}.


\section{The Delayed SEIQRPW Model with Vaccination}
\label{section:vaccine:constant}

Let us introduce in model \eqref{model}
a constant $u$ and an extra variable $W(t)$, $t\in [0, t_f ]$, 
representing the fraction of susceptible individuals that 
are vaccinated and the number of vaccines used, respectively, with 
\begin{equation} 
\label{Eq:21}
\dot{W}(t)= u\,S(t),
\end{equation}
\textls[-25]{subject to the initial condition $W(0)=0$.
Note that \eqref{Eq:21} is just the production rate of~vaccinated.}

The model with vaccination is given by the following system 
of seven nonlinear delayed differential equations:
\begin{equation} 
\label{model:vaccine}
\begin{cases}
\dot{S}(t) = b N(t) - \frac{\beta\, S(t-\tau) \,I(t-\tau)}{N(t)} 
- \left( \alpha +u+\mu\right)  \, S(t)  \, , \\[0.2 cm]
\dot{E}(t) = \frac{\beta\, S(t-\tau)\, I(t-\tau)}{N(t)} 
- (\gamma + \mu)E(t) \, , \\[0.2 cm]
\dot{I}(t) = \gamma E(t) - (\delta + \mu ) I(t) \, , \\[0.2 cm]
\dot{Q}(t) = \delta I(t) - (\lambda + \mu) Q(t)\, , \\[0.2 cm]
\dot{R}(t) = \lambda Q(t)  - \mu R(t) \, , \\[0.2 cm]
\dot{P}(t) = \alpha S(t) - \mu P(t) \, , \\[0.2 cm]
\dot{W}(t)= u\,S(t) - \mu W(t)\, , \\[0.2 cm]
\end{cases}
\end{equation}
where the total population $N(t)$ is given by
\begin{equation}
\label{totalpolpulation:N:vaccine}
N(t)= S(t)+E(t)+I(t)+Q(t)+R(t)+P(t)+W(t)\, , 
\quad \forall \, t \in [0, T] \, .
\end{equation} 
The state variables are subject to the following initial conditions: 
$S(\theta) = S_{0}$, $\theta\in [-\tau,0]$, $E(0)=E_0$, $I(\theta) = I_{0}$,
$\theta\in [-\tau,0]$, $Q(0) = Q_{0}$, $R(0)=R_0$, $P(0) = P_{0}$, and $ W(0)=0$. 

Note that in model \eqref{model:vaccine} we do not vaccinate the 
insusceptible/protected individuals $P(t)$, 
assumed protected through precautionary  measures
with a protection rate $\alpha$. Moreover, the fraction of susceptible 
individuals that are vaccinated is $u$.


\subsection{Normalized $seiqrpw$ Delayed Model with Vaccination}

Analogously to Section~\ref{sec:model1}, we consider the proportions 
of each compartment of individuals in the population, namely
$s(t)=\frac{S(t)}{N(t)}$, $e(t)=\frac{E(t)}{N(t)}$, 
$i(t)=\frac{I(t)}{N(t)}$, $q(t)=\frac{Q(t)}{N(t)}$, 
$r(t)=\frac{R(t)}{N(t)}$, $p(t)=\frac{P(t)}{N(t)}$, 
and $w(t)=\frac{W(t)}{N(t)}$. According to Equation 
\eqref{totalpolpulation:N:vaccine}, we have 
$\dot{N}(t)=(b-\mu)N(t)$. Therefore, the normalized 
$seiqrpw$ delayed model is given by
\begin{equation}
\label{model:seir:normalized:vaccine:1}
\begin{cases}
\dot{s}(t) = b- \beta\, s(t-\tau)\,i(t-\tau)- (\alpha +u+ b) \,s(t), \\[0.2 cm]
\dot{e}(t) = \beta\, s(t-\tau)\,i(t-\tau)- (\gamma + b)\,e(t),\\[0.2 cm]
\dot{i}(t) = \gamma\, e(t) - (\delta + b)\, i(t),\\[0.2 cm]
\dot{q}(t) = \delta\, i(t) - (\lambda + b) \,q(t),\\[0.2 cm]
\dot{r}(t) = \lambda \,q(t) - b \,r(t),\\[0.2 cm]
\dot{p}(t) = \alpha \,s(t) - b\, p(t),\\[0.2 cm]
\dot{w}(t) = u\,s(t) -b\,w(t) \, .
\end{cases}
\end{equation}

The state variables for system \eqref{model:seir:normalized:vaccine:1} 
are subject to the following initial conditions:   
$s(\theta) = \dfrac{S_{0}}{N(0)}$, $\theta\in [-\tau,0]$, 
$e(0)=\dfrac{E_0}{N(0)}$, $i(\theta) = \dfrac{I_{0}}{N(0)}$,
$\theta\in [-\tau,0]$, $q(0) = \dfrac{Q_{0}}{N(0)}$, 
\mbox{$r(0)=\frac{R_0}{N(0)}$, $p(0) = \dfrac{P_{0}}{N(0)}$}, 
and $w(0) = 0$, with $s(t) + e(t) + i(t) + q(t) + r(t) + p(t)+w(t) = 1$. 


\subsection{Equilibrium Points and the Basic Reproduction Number}

The disease free and the endemic equilibrium points of model 
\eqref{model:seir:normalized:vaccine:1} can be obtained by 
equating the right-hand side of Equation 
\eqref{model:seir:normalized:vaccine:1} to zero, 
hence satisfying
\begin{align*}
b- \beta\, s(t-\tau)\,i(t-\tau)- (\alpha +u+ b) \,s(t) &=0\, , \\
\beta\, s(t-\tau)\,i(t-\tau) -(\gamma+b)\,e(t)&=0\, , \\
\gamma e(t) - (\delta + b) i(t)	&=0\, , \\
\delta i(t) - (\lambda + b) q(t) &=0\, ,\\
\lambda \,q(t) - b \,r(t) &=0\, , \\
\alpha s(t) - b p(t) &=0\, , \\
u\,s(t) -b\,w(t)&=0 \, . 
\end{align*}

The disease free equilibrium of model \eqref{model:seir:normalized:vaccine:1}, 
$\Sigma_1$, is given by
\begin{equation}
\label{eq:DFE:2}
\Sigma_1 = \left(s_{0}, e_0, i_{0}, q_{0}, r_{0}, 
p_0,w_0  \right) = \left(\frac{b}{\alpha+u+b} , 
0, 0, 0, 0, \frac{\alpha}{ \alpha+u+b}, \frac{u}{ \alpha+u+b}\right) \, ,
\end{equation}
while the endemic equilibrium point for system 
\eqref{model:seir:normalized:vaccine:1}, $\Sigma_V^+$, is given by
\begin{equation}
\label{eq:VEE:vaccine}
\Sigma_V^+ = \left(s^*, e^*, i^*, q^*, r^*, p^*, w^*  \right) \, ,
\end{equation}
where\vspace{6pt}
\begin{equation*}
\begin{split}
s^* &= \frac { \left(\delta+b \right)\,\left(\gamma+b \right)}{\beta\, \gamma},\\
e^* &= \frac{\beta\,s^+\,i^+}{\left( \gamma+b \right)},\\ 
i^* &= \frac{\beta\,\gamma\,b-\left( \delta+b \right)  
\left( \gamma+b \right)\left( \alpha+u+b \right)}{\beta\,\left( \delta+b \right)
\, \left( \gamma+b \right)},\\
q^* &=\frac{\beta\,\gamma\,b\, \delta-\delta\,\left( \delta+b \right)  
\left( \gamma+b \right)  \left( \alpha+u+b \right)}{\beta\,
\left( \lambda+b \right)\,\left( \delta+b \right)\,  \left( \gamma+b \right)},\\
r^* &= \frac{\lambda\,\delta\,\beta\,\gamma\,b-\lambda\,
\delta\,\left( \delta+b \right)  \left( \gamma+b \right)  
\left( \alpha+u+b \right)}{b\,\beta\,\left( \lambda+b \right)
\,\left( \delta+b \right)\,  \left( \gamma+b \right)},\\ 
p^* &= \frac { \alpha\,\left(\delta+b \right)\,
\left(\gamma+b \right)}{b\,\beta\, \gamma},\\ 
w^* &= \frac { u\, \left(\delta+b \right)
\,\left(\gamma+b \right)}{b\, \beta\, \gamma}.
\end{split}
\end{equation*}

Following the method from van den Driessche \cite{MR1950747}, 
we obtain the following basic reproduction number, 
denoted by $\tilde{R}_0$:
\begin{equation}
\label{eq:R0:vaccine}
\tilde{R}_0 = \frac {\beta\,\gamma\,b}{\left( \alpha+u+b \right) 
\left( \delta+b \right)  \left( \gamma+ b \right)   } \, .
\end{equation}


\subsection{Stability of the Normalized $seiqrpw$ Delayed Model with Vaccination}
\label{subsec:stability:vac}

Consider the following coordinate transformation:
$x_1(t)=s(t)-\bar{s}$, $x_2(t)=e(t)-\bar{e}$, 
$x_3(t)=i(t)-\bar{i}$, $x_4(t)=q(t)-\bar{q}$, 
$x_5(t)=r(t)-\bar{r}$, $x_6(t)=p(t)-\bar{p}$, 
and $x_7(t)=w(t)-\bar{w}$,
where $(\bar{s},\bar{r},\bar{i},\bar{q},\bar{r},\bar{p},\bar{w})$ 
denotes an equilibrium point of system \eqref{model:seir:normalized:vaccine:1}. 
The linearized system of \eqref{model:seir:normalized:vaccine:1} takes the form
\begin{equation}
\dot{X}(t)=\tilde{A}_0\,X(t)+\tilde{A}_1\,X(t-\tau),
\end{equation} 
where $X=(x_1,x_2,x_3,x_4,x_5,x_6,x_7)^T$,
\begin{equation*}
\tilde{A}_0=\begin{pmatrix}
-\alpha-u-b & 0 & 0 & 0 & 0 & 0& 0\\
0 & -\gamma-b & 0 & 0 & 0 & 0& 0 \\
0 & \gamma & -\delta-b & 0 & 0 & 0& 0\\ 
0 & 0 & \delta & -\lambda-b & 0 & 0& 0\\
0 & 0 & 0 & \lambda & -b &0& 0  \\
\alpha & 0 & 0 & 0 & 0 &-b& 0 \\
u & 0 & 0 & 0 & 0 &0& -b \\
\end{pmatrix}, 
\end{equation*}
\begin{equation*}
\tilde{A}_1
=\begin{pmatrix}
-\beta\,\bar{i} & 0 & -\beta\,\bar{s} & 0 & 0 & 0& 0\\
\beta\,\bar{i} & 0 &  \beta\,\bar{s} & 0 & 0 &0& 0\\
0 & 0 & 0 & 0 & 0 & 0& 0 \\
0 & 0 & 0 & 0 & 0 & 0 & 0\\
0 & 0 & 0 & 0 & 0 & 0& 0 \\
0 & 0 & 0 & 0 & 0 & 0 & 0\\
0 & 0 & 0 & 0 & 0 & 0 & 0
\end{pmatrix}. 
\end{equation*}

The characteristic equation of system 
\eqref{model:seir:normalized:vaccine:1} is given by
\begin{equation}
\label{caract:vaccine}
\tilde{\Gamma}(y)=\vert y\, Id_{7\times 7}
-\tilde{A}_0-\tilde{A}_1\,e^{-\tau\,y}\vert.
\end{equation}

We are also able to prove stability results 
for the normalized $seiqrpw$ delayed 
model with vaccination.

\begin{Theorem}[Stability of the disease free equilibrium of system \eqref{model:seir:normalized:vaccine:1}]
\label{theo3}
If $\tilde{R}_0<1$, then the disease free equilibrium $\Sigma_1$ is locally 
asymptotically stable for any time-delay $\tau\geq 0$. If  $\tilde{R}_0>1$, 
then the disease free equilibrium is unstable for any time-delay $\tau\geq 0$. 
\end{Theorem}

\begin{proof}
The characteristic Equation \eqref{caract:vaccine} at the disease free equilibrium, 
$\Sigma_1$, is given by
\begin{equation}
\label{characteristic:equation:vaccine}
P^*(y,\tau)=(y+b)^3\,(y+\alpha+u+b)\,(y+\lambda+b)\,(y^2+\Gamma _1\,y+\Gamma_2(y))=0,
\end{equation}
where $\Gamma_1=\delta+2\,b+\gamma$ and 
$\Gamma_2(y)=(\delta+b)\,(\gamma+b)
-\dfrac{\beta\,\gamma\,b}{\alpha+u+b }e^{-\tau \,y}$.
\begin{itemize}
\item[(i)] Let $\tau=0$. In this case, the Equation 
\eqref{characteristic:equation:vaccine} becomes
\begin{multline}
\label{characteristic:equation:tau0:v}
P^*(y,0)=(y+b)^3\,(y+\alpha+u+b)\,(y+\lambda+b)\\
\left( y^2+\Gamma_1\,y+(\delta+b)\,(\gamma+b)-\dfrac{\beta\,\gamma\,b}{\alpha+u+b }\right) =0 \, .
\end{multline}
We need to prove that all roots of the characteristic Equation \eqref{characteristic:equation:tau0:v} 
have negative real parts. It is easy to see that $y_1=-b$, $y_2=-\alpha-u-b$ and $y_3=-\lambda-b$ 
are roots of Equation \eqref{characteristic:equation:tau0:v} and the three are real and negative. 
Thus, we just need to consider the fourth term of Equation \eqref{characteristic:equation:tau0:v}. 
Let
\begin{equation*}
P_3^*(y,0):=y^2+\Gamma_1\,y+(\delta+b)\,(\gamma+b)-\dfrac{\beta\,\gamma\,b}{\alpha+u+b}.
\end{equation*}
Using the Routh--Hurwitz criterion \cite{Rogers:Routh:Hurwitz}, 
we know that all roots of $P_3^*(y,0)$ 
have negative real parts if, and only if, the coefficients of $P_3^*(y,0)$ 
are strictly positive. In this case, $\Gamma_1=\delta+2\,b+\gamma >0$ and 
$$
(\delta+b)\,(\gamma+b)-\dfrac{\beta\,\gamma\,b}{\alpha+u+b } >0 \,  
\textrm{ if, and only if,}\, \, 
\tilde{R}_0= \frac {\beta\,\gamma\,b}{ \left( \alpha+u+b \right)\, 
\left( \delta+b \right) \, \left( \gamma + b \right)   }<1.
$$
Therefore, we have proved that the disease free equilibrium, $\Sigma_1$, 
is locally asymptotically stable for $\tau=0$, whenever $ \tilde{R}_0 < 1$.
		
\item[(ii)] Let $\tau>0$. Using Rouch\'e's theorem, we prove that all roots 
of the characteristic Equation \eqref{characteristic:equation:vaccine} cannot 
have pure imaginary roots. Suppose the contrary, i.e., that there exists 
$w\in \mathbb{R}$ such that $y = w\,i$ is a solution 
of \eqref{characteristic:equation:vaccine}.
Replacing $y$ in the fourth term of~\eqref{characteristic:equation:vaccine}, we get
\begin{equation*}
-w^2+(\delta+2\,b+\gamma)\,w\,i+(\delta+b)\,(\gamma+b)
-\dfrac{\beta\,\gamma\, b}{\alpha+u+b } 
\left(\cos (\tau\,w)-i\,\sin (\tau\,w)\right)=0 \, .
\end{equation*}
Then,
\begin{equation*}
\begin{cases}
-w^2+(\delta+b)\,(\gamma+b) 
= \dfrac{\beta\, \gamma\,b}{\alpha+u+b}\,\cos(\tau\,w),\\[0.2 cm]
(\delta+2\,b+\gamma)\,w =-\dfrac{\beta\,\gamma\,b}{\alpha+u+b}
\,\sin (\tau\,w) \, .
\end{cases}
\end{equation*}
By adding up the squares of both equations and using 
the fundamental trigonometric formula, one has
\begin{equation*}
w^4+\left( \left( \delta+b\right)^2
+\left( \gamma+b\right) ^2\right) w^2+
(\delta+b)^2\,(\gamma+b)^2
-\left(\dfrac{\beta\,\gamma\,b}{\alpha+u +b}\right)^2=0,
\end{equation*} 
which is equivalent to
\begin{equation}
w^2= \dfrac{1}{2} \sqrt{\left( \left( \delta+b\right)^2
-\left( \gamma+b\right) ^2\right)^2  
+4 \,\left(\dfrac{\beta\,\gamma\,b}{\alpha+u +b}\right)^2}
-\dfrac{1}{2}\left( \left( \delta+b\right)^2
+\left( \gamma+b\right) ^2\right). 
\end{equation}
If $\tilde{R}_0<1$, then 
$(\delta+b)^2\,(\gamma+b)^2
-\left(\dfrac{\beta\,\gamma\,b}{\alpha+u +b}\right)^2 >0$, and 
\begin{small}	
\begin{equation*}
\left( \left( \delta+b\right)^2
+\left( \gamma+b\right)^2\right)^2
-4\left( (\delta+b)^2\,(\gamma+b)^2
-\left(\dfrac{\beta\,\gamma\,b}{\alpha+u +b}\right)^2 \right)
<\left( \left( \delta+b\right) ^2+\left( \gamma+b\right) ^2\right)^2,
\end{equation*}
\end{small}
so that
$$
\sqrt{\left( \left( \delta+b\right)^2-\left( \gamma+b\right) ^2\right)^2  
+4 \,\left(\dfrac{\beta\,\gamma\,b}{\alpha+u +b}\right)^2}
<\left( \delta+b\right) ^2+\left( \gamma+b\right) ^2.
$$
Hence, we have $w^2<0,$ which is a contradiction. Therefore, we have proved that 
if $\tilde{R}_0 < 1$, then the characteristic Equation 
\eqref{characteristic:equation:vaccine} cannot have pure imaginary roots 
and the disease free equilibrium $\Sigma_1$ is locally asymptotically stable, 
for any strictly positive time delay $\tau$.
		
\item[(iii)] Suppose now that $\tilde{R}_0 > 1$. 
We know that the characteristic Equation 
\eqref{characteristic:equation:vaccine} has three real negative 
roots $y_1=-b, y_2 =-\alpha-u-b$ and $y_3=-\lambda-b.$ Thus, 
we need to check if the remaining roots of
\begin{equation}
q^*(y):=y^2+\Gamma_1\,y+\Gamma_2(y)
\end{equation}
have negative real parts. It is easy to see that 
$q(0)=\Gamma_2(0)<0$, because we are
assuming $\tilde{R}_0 > 1$. On the other hand,
$\lim\limits_{y\rightarrow +\infty} q^*(y)=+\infty$. Therefore, by continuity 
of $q^*(y)$, there is at least one positive root of the characteristic Equation 
\eqref{characteristic:equation:vaccine}. Hence, we conclude 
that $\Sigma_1$ is unstable, for any $\tau \geq 0$.
\end{itemize}	
The proof is complete.
\end{proof}

\begin{Theorem}[Stability of the endemic equilibrium point of system \eqref{model:seir:normalized:vaccine:1}]
\label{theo4}
Let $\tau= 0$. If $\tilde{R}_0>1$, then the endemic equilibrium point 
$\Sigma_V^+$ is locally asymptotically stable. When $\tau > 0$, 
the endemic equilibrium point $\Sigma_V^+$ is locally asymptotically stable 
if the basic reproduction number $\tilde{R}_0$ satisfies the following relations:
\begin{equation}
\label{tilde:R0:condition:1}
1<\tilde{R}_0<\min{ \left( 3,1+\dfrac{\sqrt{(\alpha+u+b)^2
+(\delta+b)^2+(\gamma+b)^2}}{\alpha+u+b} \right)} 
\end{equation}
and
\begin{equation}
\label{tilde:R0:condition:2}
M_1^* \tilde{R}_0^2+M_2^* \tilde{R}_0+M_3^*>0 \, ,
\end{equation}
where
\begin{equation}
\begin{split}
M_1^*&=-(\alpha+u+b)^2\,\left((\delta+b)^2+(\gamma+b)^2
\right),\\ 
M_2^*&=2\,\left( \alpha+u+b\right)^2\left(\left( \delta+\gamma
+2\, b\right) ^2-3\,(\delta+b)\,(\gamma+b)\right)\\  
&\qquad +2\,\left( \alpha+u+b\right) \,(\delta+b)\,(\gamma+b)\,(\delta+\gamma+2\, b),\\
M_3^*&=2\,(\alpha+u+b)\,(\delta+b)\,(\gamma+b)\left( \alpha+u-\delta-\gamma-b  \right). 
\end{split}
\end{equation}
\end{Theorem}

\begin{proof}
\textls[25]{The characteristic Equation \eqref{caract:vaccine}, computed 
at the endemic equilibrium $\Sigma_V^+$, is given~by}
\begin{equation}
\label{characteristic:equation:EE:V}
\tilde{P}^*(y,\tau)=(y+b)^3\,(y+\lambda+b)
\,(y^3 +\Omega_1(y)\, y^2+\Omega_2(y)\,y+\Omega_3(y))=0 \, , 
\end{equation}
where
$\Omega_1(y)=L_1^*+\bar{L}_1^*\,e^{-\tau\,y}$,
$\Omega_2(y)=L_2^*+\bar{L}_2^*\,e^{-\tau\,y}$, 
and $\Omega_3(y)=L_3^*+\bar{L}_3^*\,e^{-\tau\,y}$
with \vspace{6pt}
\begin{equation*}
\begin{split}
L_1^* &= \alpha+u+\delta+\gamma+3\,b,\\ 
\bar{L}_1^* &=	\frac{\beta\,\gamma\,b-\left( \delta+b \right)  
\left( \gamma+b \right)  \left( \alpha+u+b \right)}{\left( \delta+b \right)
\, \left(\gamma+b\right)},\\ 
L_2^* &=(\gamma+2\,b+\delta)\,(\alpha+u+b)
+(\gamma+b)\,(\delta+b),\\ 
\bar{L}_2^* &= (\gamma+2\,b+\delta)
\,(\alpha+u+b)\,(\tilde{R}_0-1)-(\gamma+b)\,(\delta+b),\\ 
L_3^* &=(\alpha+u+b)\,(\gamma+b)\,(\delta+b),\\  
\bar{L}_3^*&=\beta\,\gamma\,b-2\,(\alpha+u+b)\,(\gamma+b)\,(\delta+b).
\end{split}
\end{equation*}
\begin{itemize}
\item[(i)]
Let $\tau=0$. In this case, Equation \eqref{characteristic:equation:EE:V} becomes
\begin{equation}
\label{characteristic:equation:tau0:EE:V}
\tilde{P}^*(y,0)=(y+b)^3\,\left( y+\lambda+b\right) 
\,\left( y^3 +\tilde{\Omega}_1\, y^2+\tilde{\Omega}_2\,y+\tilde{\Omega}_3\right) =0 \, ,
\end{equation}
where $\tilde{\Omega}_1=L_1^*+\bar{L}_1^*\,$,
$\tilde{\Omega}_2=L_2^*+\bar{L}_2^*\,$ and $\tilde{\Omega}_3=L_3^*+\bar{L}_3^*$.
Looking at the roots of the characteristic Equation \eqref{characteristic:equation:tau0:EE:V}, 
it is easy to see that $y_1=-b$ and $y_2=-\lambda-b$ are real negative roots of 
\eqref{characteristic:equation:tau0:EE:V}. Considering the third term of the above equation, let
\begin{equation}
\label{P:tilde:star:vac}
\tilde{P}_3^*(y,0):= y^3 +\tilde{\Omega}_1\, y^2+\tilde{\Omega}_2\,y+\tilde{\Omega}_3=0	 \, .
\end{equation}
Using the Routh--Hurwitz criterion \cite{Rogers:Routh:Hurwitz}, 
we know that all roots of $\tilde{P}_3^*(y,0)$ have negative real parts 
if, and only if, the coefficients of $\tilde{P}_3^*(y,0)$ 
are strictly positive and 
$$
\tilde{\Omega}^*=\tilde{\Omega}_1\,\tilde{\Omega}_2-\tilde{\Omega}_3\,>0.
$$ 
If $\tilde{R}_0>1$, then
\begin{equation*}
\begin{split}
\tilde{\Omega}_1&=\alpha+u+\delta+\gamma+3\,b+\left( \alpha+u+b\right) 
\,(\tilde{R}_0-1) >0,\\
\tilde{\Omega}_2&=(\delta+\gamma+2\,b)\,(\alpha+u+b)\, \tilde{R}_0 >0,\\
\tilde{\Omega}_3&=(\alpha+u+b)\,(\delta+b)\,(\gamma+b)\,(\tilde{R}_0-1)>0,\\
\tilde{\Omega}^*&=(\alpha+u+b)\,(\delta+\gamma+2\,b)\,\tilde{R}_0^2
+(\alpha+u+b)\,(\delta^2+3\,b\,(\delta+b)+\gamma\,(\delta+\gamma+3\,b))
\,\tilde{R}_0\\
&\quad + (\alpha+u+b)\, (\delta+b)\,(\gamma+b) >0.
\end{split}
\end{equation*}
		
\item[(ii)] \textls[-25]{Let $\tau>0$. By Rouch\'e's theorem, we prove that all roots 
of the characteristic \mbox{Equation~\eqref{characteristic:equation:EE:V}} }
cannot intersect the imaginary axis, i.e., the characteristic equation 
cannot have pure imaginary roots. Suppose the opposite, i.e., that there 
exists $w\in \mathbb{R}$ such that $y = w\,i$ is a solution of 
\eqref{characteristic:equation:EE:V}. Replacing $y$ in the third term 
of \eqref{characteristic:equation:EE:V}, we get
\begin{equation*}
-w^3\,i-L_1^*\,w^2+L_2^*\,w\,i+L_3^* + (-\bar{L}_1^*\,w^2+\bar{L}_2^*\,w\,i+\bar{L}_3^*)
\,\left(\cos (\tau\,w)-i\,\sin (\tau\,w)\right)=0.
\end{equation*}
Then,
\begin{equation*}
\begin{cases}
-L_1^*\,w^2+L_3^*=(\bar{L}_1^*\,w^2-\bar{L}_3^*)\,
\cos(\tau\,w)-\bar{L}_2^*\,w\,\sin (\tau\,w)\, ,\\[0.2 cm]
-w^3+L_2^*\,w =- \bar{L}_2^*\,w \,\cos(\tau\,w)
- (\bar{L}_1^*\,w^2-\bar{L}_3^*)\,\sin (\tau\,w) \, .
\end{cases}
\end{equation*}
By adding up the squares of both equations and using 
the fundamental trigonometric formula, we obtain that
\begin{equation*}
w^6+K_1^*\,w^4+K_2^*\,w^2+K_3^*=0,
\end{equation*}
where
\begin{equation*}
\begin{split}
K_1^*&=(L_1^*)^2-(\bar{L}_1^*)^2-2\,L_2^*,\\
K_2^*&=2\,\bar{L}_1^*\,\bar{L}_3^*-2\,L_1^*\,L_3^*+(L_2^*)^2-(\bar{L}_2^*)^2,\\ 
K_3^*&=(L_3^*)^2-(\bar{L}_3^*)^2. 
\end{split}
\end{equation*}

Assume that the basic reproduction number $\tilde{R}_0$ satisfies relations 
\eqref{tilde:R0:condition:1} and \eqref{tilde:R0:condition:2} with the condition
\begin{multline}
\label{condition:2:V}
\min{ \left( 3,1+\dfrac{\sqrt{(\alpha+u+b)^2
+(\delta+b)^2+(\gamma+b)^2}}{\alpha+b} \right)}\\
=1+\dfrac{\sqrt{(\alpha+u+b)^2+(\delta+b)^2
+(\gamma+b)^2}}{\alpha+u+b}.
\end{multline}
Then,
\begin{equation*}
K_1^*=(\delta+b)^2+(\gamma+b)^2+(\alpha+u+b)^2\,
\left(1-\left( \tilde{R}_0-1\right)^2 \right)>0.
\end{equation*}
In contrast, if $\tilde{R}_0$ satisfies relations 
\eqref{tilde:R0:condition:1} and \eqref{tilde:R0:condition:2}
under the condition
\begin{equation}
\label{condition:1:V}
\min{ \left( 3,1+\dfrac{\sqrt{(\alpha+u+b)^2
+(\delta+b)^2+(\gamma+b)^2}}{\alpha+u+b} \right)}=3,
\end{equation}
then we have	
\begin{equation*}
1<\tilde{R}_0<3< 1+\dfrac{\sqrt{(\alpha+u+b)^2
+(\delta+b)^2+(\gamma+b)^2}}{\alpha+u+b},
\end{equation*}
which is equivalent to
\begin{equation*}
\begin{split}
&0<\tilde{R}_0-1<2<\dfrac{\sqrt{(\alpha+u+b)^2
+(\delta+b)^2+(\gamma+b)^2}}{\alpha+u+b},\\
&1-\left( \dfrac{(\alpha+u+b)^2+(\delta+b)^2
+(\gamma+b)^2}{(\alpha+u+b)^2}\right) <1-(\tilde{R}_0-1)^2<1,\\
&-(\delta+b)^2-(\gamma+b)^2<(\alpha+u+b)^2\left( 1
-(\tilde{R}_0-1)^2\right) <(\alpha+u+b)^2.
\end{split}
\end{equation*}
Thus,
\begin{equation*}
K_1^*>0.
\end{equation*}
Under the assumption that the basic reproduction number $\tilde{R}_0$ 
satisfies relations \eqref{tilde:R0:condition:1} 
and \eqref{tilde:R0:condition:2}, we have
\begin{equation}
K_2^*=M_1^* \tilde{R}_0^2+M_2^* \tilde{R}_0+M_3^*>0 \, .
\end{equation}
Therefore, if we assume that the basic reproduction number 
$\tilde{R}_0$ satisfies relations~\eqref{tilde:R0:condition:1} 
and \eqref{tilde:R0:condition:2} with condition \eqref{condition:1:V}, then
\begin{equation}
K_3^*=(\alpha+u+b)^2\,(\delta+b)^2\,(\gamma+b)^2
\,\left(1-\left(\tilde{R}_0-2\right)^2 \right)>0;
\end{equation}
if $\tilde{R}_0$ satisfies \eqref{tilde:R0:condition:1} 
and \eqref{tilde:R0:condition:2} with condition 
\eqref{condition:2:V}, then we have
\begin{equation*}
1<\tilde{R}_0<	1+\dfrac{\sqrt{(\alpha+u+b)^2
+(\delta+b)^2+(\gamma+b)^2}}{\alpha+u+b}<3 \, ,
\end{equation*}
which is equivalent to
\begin{equation*}
-1<\tilde{R}_0-2 < -1+\dfrac{\sqrt{(\alpha+u+b)^2
+(\delta+b)^2+(\gamma+b)^2}}{\alpha+u+b}<1,
\end{equation*}
and also equivalent to
\begin{equation*}
1-\left( \tilde{R}_0-2\right)^2>0 \, .
\end{equation*}
Thus,
\begin{equation*}
K_3^*>0.
\end{equation*}
We have just proved that the left hand-side of Equation \eqref{characteristic:equation:EE:V} 
is strictly positive, which implies that this equation is not possible. Therefore, 
\eqref{characteristic:equation:tau0:EE:V} does not have imaginary roots, and $\Sigma_V^+$ 
is locally asymptotically stable, for any time delay $\tau > 0$, whenever $\tilde{R}_0$ 
satisfies conditions \eqref{tilde:R0:condition:1} and \eqref{tilde:R0:condition:2}.
\end{itemize}	
The proof is complete.
\end{proof}

It should be noted that Theorem~\ref{theo3} is not trivial, 
and it is not easy to give a biological/medical interpretation 
to the relations \eqref{tilde:R0:condition:1} 
and \eqref{tilde:R0:condition:2}.


\section{Numerical Simulations and Discussion}
\label{sec:numerical}

In this section we investigate, numerically, the local stability 
of the normalized $seiqrp$ and $seiqrpw$ models, 
illustrating our results from Sections~\ref{sec:model1} 
and \ref{section:vaccine:constant}. 
All numerical computations were performed in the 
numeric computing environment \textsf{MATLAB R2019b} 
using the medium order method and numerical 
interpolation \cite{MR1433374}.


\subsection{Local Stability of the Delayed $seiqrp$ Model}
\label{sec:example:01}

\begin{figure}[H]
\includegraphics[scale=0.6]{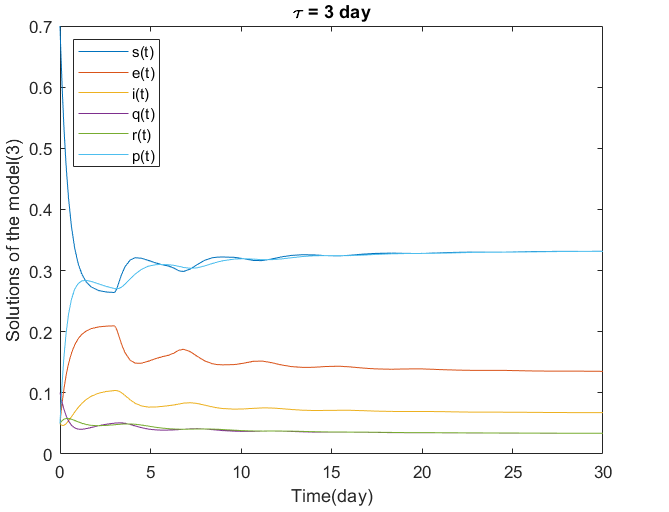}
\caption{Dynamics of model \eqref{model:seir:normalized} with $\tau=3$ days. 
Parameter values as in Table~\ref{table:parameter:values:0}.}
\label{example:01}
\end{figure}

Consider the normalized delayed $seiqrp$ model \eqref{model:seir:normalized}, 
proposed in Section~\ref{sec:model1}. Take the initial conditions 
\begin{equation*}
(s_0, e_0, i_0, q_0, r_0, p_0)
=\left( 0.7, 0.05,0.05,0.1, 0.05,0.05\right)
\end{equation*}
and the parameter values as given 
in Table~\ref{table:parameter:values:0}.
\begin{table}[H]
\caption{Parameter values used in the simulations 
of Section~\ref{sec:example:01}.}	
\label{table:parameter:values:0}
\setlength{\tabcolsep}{11.1mm}
\begin{tabular}{c c c c} \toprule
\textbf{Parameter} & \textbf{Value} & \textbf{Units} & \textbf{Ref} \\ \midrule
$b$       & 1  &              & Assumed  \\ 
$\mu$     & 1  &              & Assumed  \\
$\delta$ &  1 &    & Assumed   \\
$\alpha$  & 1  &   day$^{-1}$ & Assumed   \\
$\beta$   & 12 &  day$^{-1}$  & Assumed   \\
$\gamma$  &  1 &   day$^{-1}$ & Assumed   \\
$\lambda$ & 1  &  day$^{-1}$  & Assumed   \\ 
$t_f$     & 30 &  day         & Assumed  \\ \bottomrule
\end{tabular}
\end{table}
In Figure~\ref{example:01}, we present the numerical solutions 
to the delayed model \eqref{model:seir:normalized} 
in the time interval $[0, 30]$ days.

Considering the parameter values from Table~\ref{table:parameter:values:0}, 
we have the following value for the basic reproduction number $R_0$ 
of Section \ref{sec:model1}: $R_0=1.5$.
From Theorem~\ref{theo2}, $R_0=1.5$ satisfies the conditions \eqref{R0:condition:1} 
and \eqref{R0:condition:2}, so the endemic equilibrium point 
$EE=(\frac{1}{3}, \frac{1}{6},\frac{1}{12},\frac{1}{24},\frac{1}{24},\frac{1}{3})$ 
of system \eqref{model:seir:normalized} is locally asymptotically stable 
for any time delay $\tau\geq 0$.

\begin{figure}[H]
\includegraphics[scale=0.5]{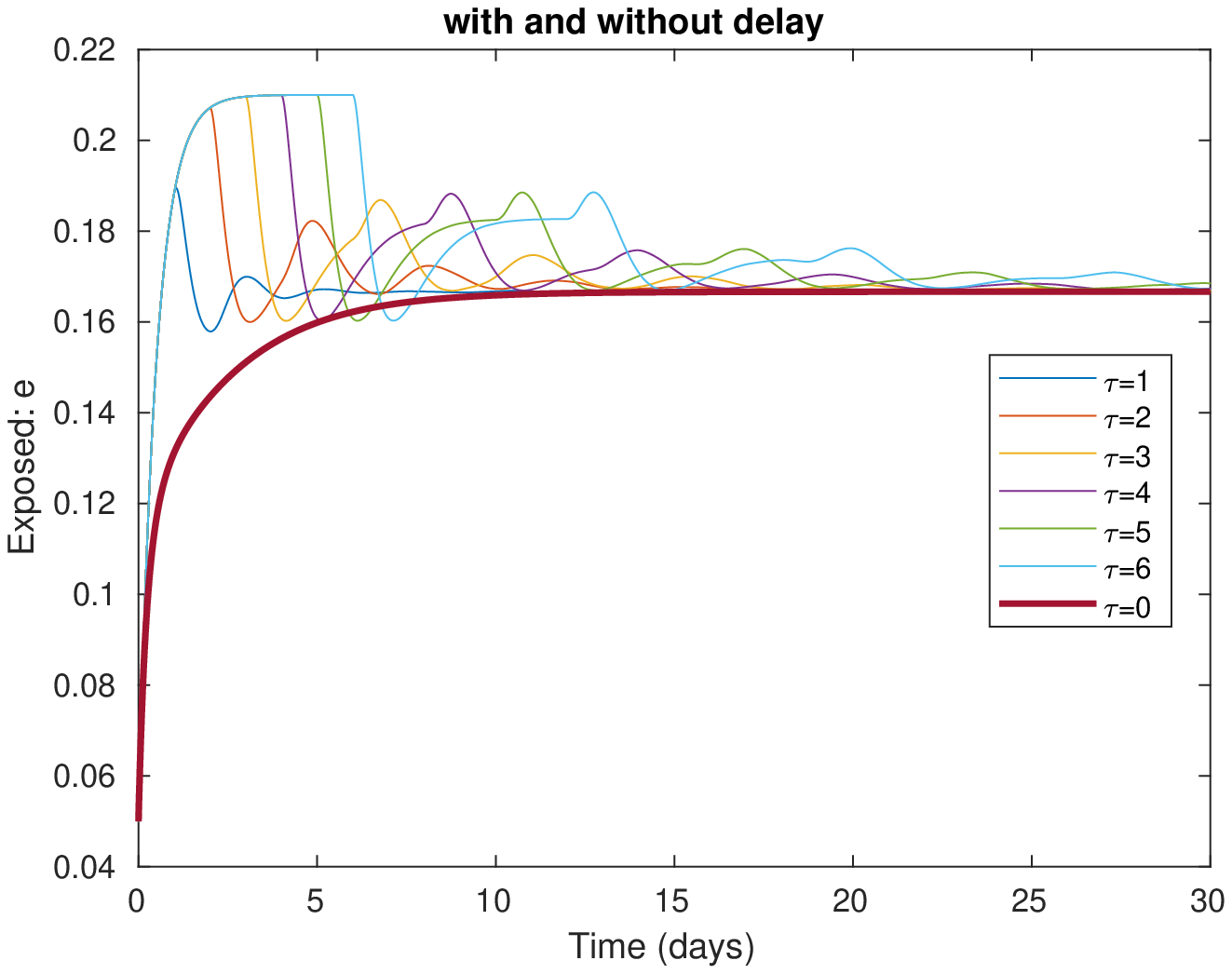}
\includegraphics[scale=0.5]{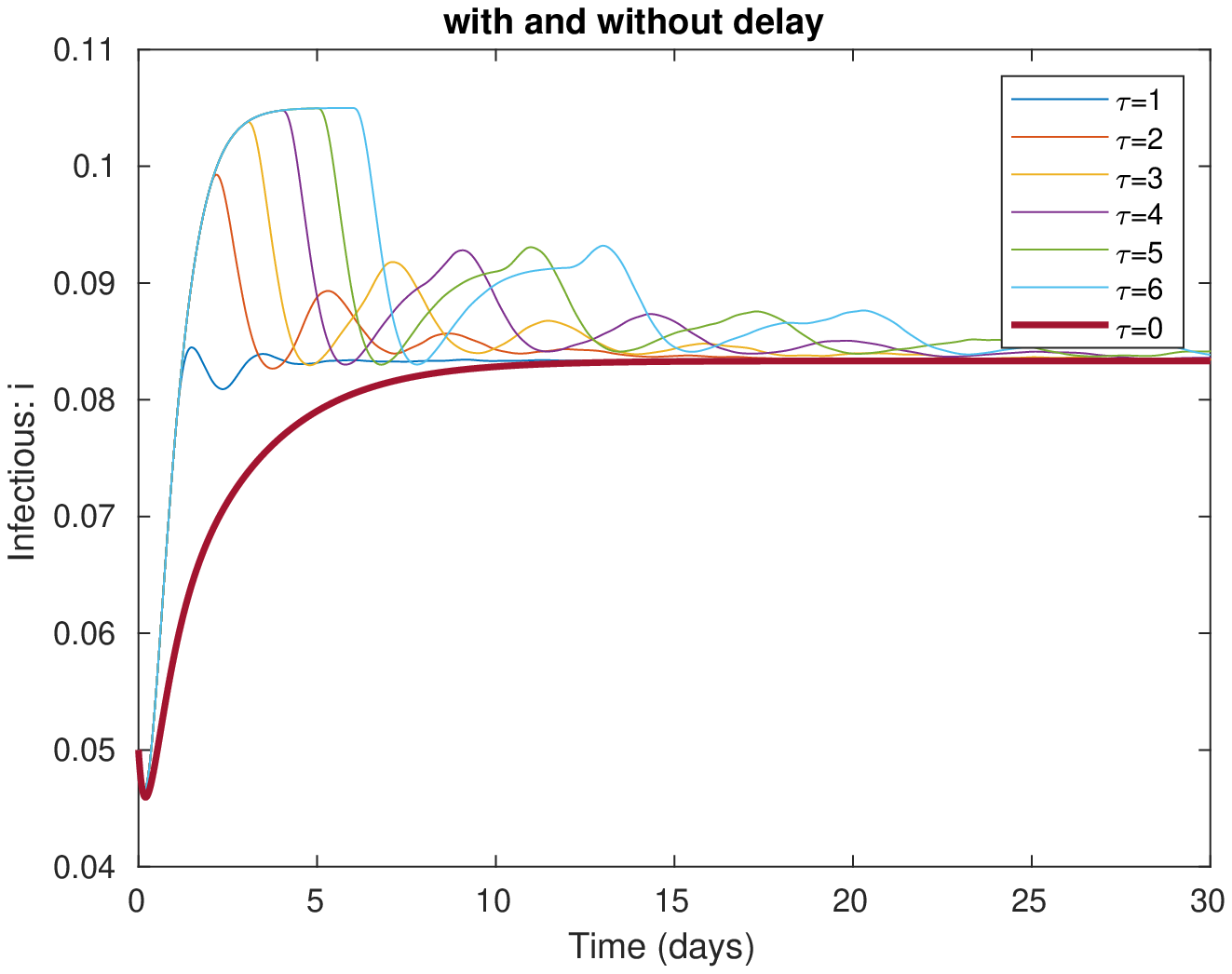}
\caption{Dynamics of model \eqref{model:seir:normalized} with $\tau \in [0, 6]$ days.
Parameter values as in Table~\ref{table:parameter:values:0}.}
\label{example:01:2}
\end{figure}

In Figure~\ref{example:01:2}, we observe the effect of the time delays: 
$\tau = 0, \ldots, 6$ on the classes $e$ of exposed and $i$ of infectious. 
The presence of waves is due to the presence of the time delay
and is related to the emergence of the COVID-19 waves.
For the study of multiple epidemic waves in the context of COVID-19, 
we refer the interested reader to \cite{MR4441440}.


\subsection{Delayed $seiqrpw$ Model with Vaccination: COVID-19 in Italy}
\label{sec:example:02}

Now, we study, numerically, the stability of the spread of the epidemic of COVID-19 in Italy 
for the period of three months starting from 18 October  2020, using the delayed
model~\eqref{model:seir:normalized:vaccine:1} that we proposed 
in Section~\ref{section:vaccine:constant}.~The preliminary conditions 
and real data were taken and computed from the database
\url{https://raw.githubusercontent.com/pcm-dpc/COVID-19/master/dati-regioni/dpc-covid19-ita-regioni.csv}
(accessed on 14 August 2021). We consider the initial \mbox{conditions} 
\begin{equation*}
(s_0, e_0, i_0, q_0, r_0, p_0,w_0)
=\frac{1}{N}\left( 59.769.273,403.601,8.837,44.098,254.058,133,0\right)
\end{equation*}
with $N=60.480.000$ \cite{United:Nations}, and the parameter values as given 
in Table~\ref{table:parameter:values}, which are obtained 
using the nonlinear least-squares solver \cite{Cheynet}.
The reader interested in the details of the nonlinear least-squares solver,
according to which the parameters of the delayed model~\eqref{model:seir:normalized:vaccine:1} 
are computed, is referred to the open access 
article \cite{Cheynet}.

\vspace{-6pt}

\vspace{-6pt}

\begin{table}[H]
\caption{Parameter values used in the simulations 
of Section~\ref{sec:example:02}, modeling the spread 
of the epidemic of COVID-19 in Italy for the period 
of three months starting 18 October 2020.}	
\label{table:parameter:values}
\setlength{\tabcolsep}{10.3mm}
\begin{tabular}{c c c c} \toprule
\textbf{Parameter} & \textbf{Value}             & \textbf{Units}      &  \textbf{Ref}\\ \midrule
$b$       & 7.391\textperthousand  &            & \cite{United:Nations} \\ 
$\mu$     & 10.658\textperthousand &            & \cite{United:Nations} \\
$\alpha$  & 1.1775               &  day$^{-1}$  & \cite{Cheynet}  \\
$\beta$   & 3.97                 &  day$^{-1}$  & \cite{Cheynet} \\
$\gamma$  &  0.0048              &  day$^{-1}$  & \cite{Cheynet}  \\
$\lambda$ & 0.0182256            &  day$^{-1}$  & \cite{Cheynet}   \\ 
$\delta$  & 0.1432   &              &  \cite{Cheynet}   \\
$t_f$      & 90                  & day          & Assumed \\ \bottomrule
\end{tabular}
\end{table}

\begin{figure}[H]
\begin{center}
\includegraphics[scale=0.35]{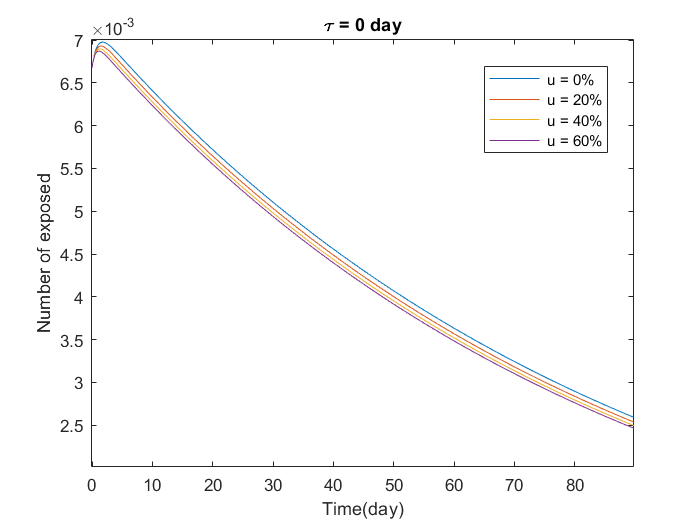}
\end{center}
\includegraphics[scale=0.35]{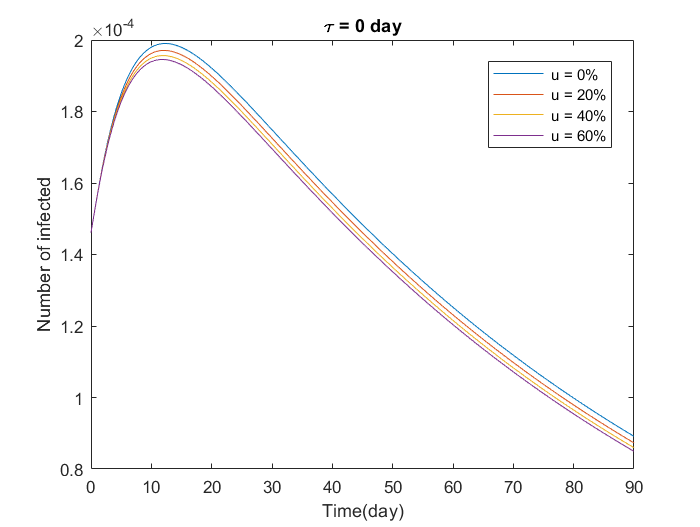}
\includegraphics[scale=0.35]{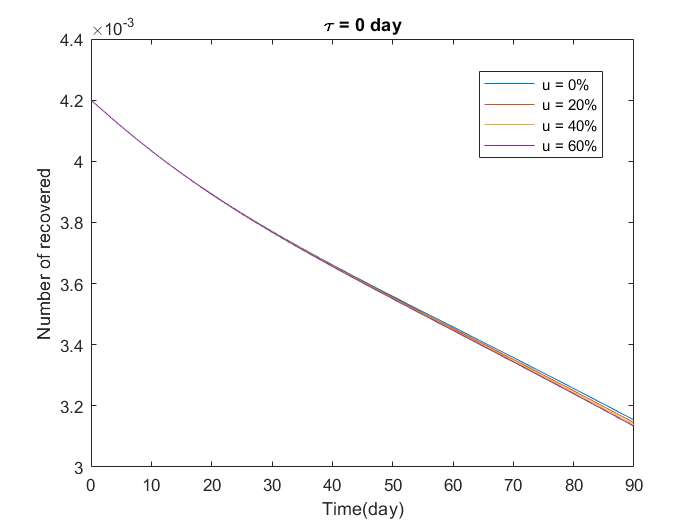}
\caption{Predictions for Italy from the delayed model 
\eqref{model:seir:normalized:vaccine:1} 
with $\tau=0$ and $u\in\{0\%, 20\%, 40\%, 60\%\}$,
between 18 October 2020, and 19 January 2021.}
\label{fig1:tau:fixe}
\end{figure}

\begin{figure}[H]
\begin{center}
\includegraphics[scale=0.35]{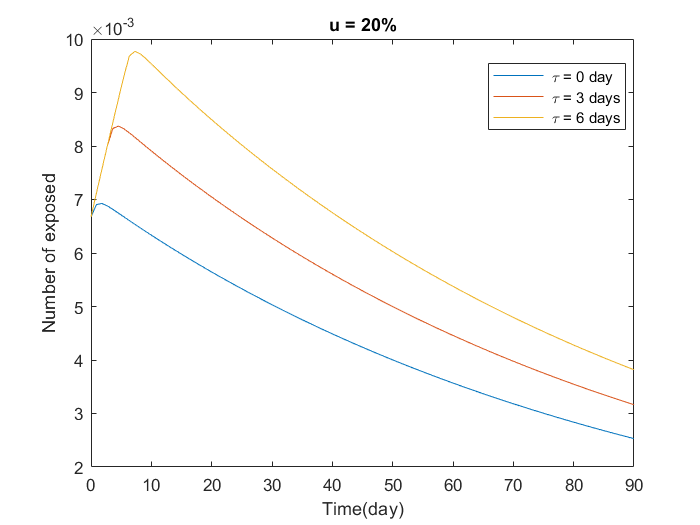}
\end{center}
\includegraphics[scale=0.35]{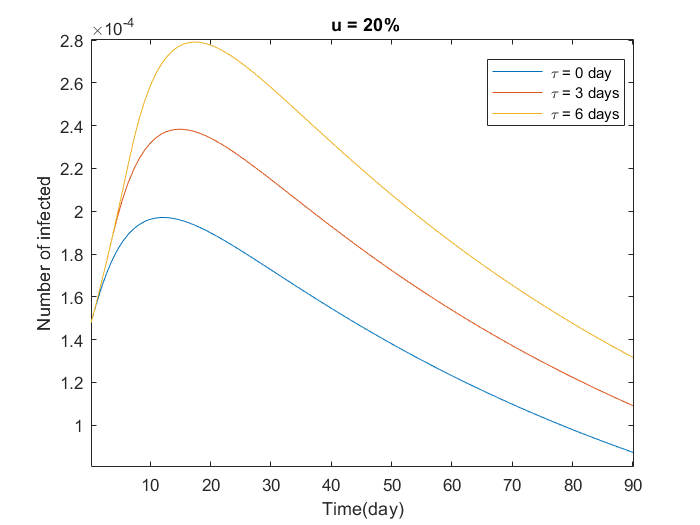}
\includegraphics[scale=0.35]{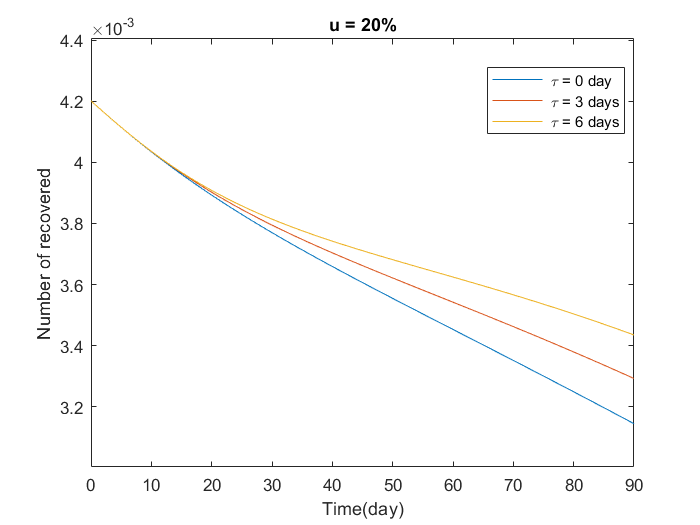}
\caption{Predictions for Italy from the delayed model 
\eqref{model:seir:normalized:vaccine:1} 
with $u=20\%$ and $\tau\in\{0, 3 , 6 \}$ days, 
between 18 October 2020, and 19 January 2021.}
\label{fig1:u:fixe}
\end{figure}

In Figures~\ref{fig1:tau:fixe} and \ref{fig1:u:fixe}, we present 
numerical solutions to the delayed model \eqref{model:seir:normalized:vaccine:1} 
in the time interval $t \in [0, 90]$ days, $t = 0$ representing 18 October 2020,
and considering two~cases:
\begin{itemize}
\item Case~1: $\tau=0$ days (without delay), with different percentages 
of susceptible individuals being vaccinated --- $u=0\%$, $u=20\%$, $u=40\%$ 
and $u=60\%$ (Figure~\ref{fig1:tau:fixe}).

\item Case~2: $u=20\%$ (fixed), with different delays --- $\tau=0$ days, $\tau=3$ days, 
and $\tau=6$ days (Figure~\ref{fig1:u:fixe}).
\end{itemize}
\vspace{-6pt}

Considering the parameter values from Table~\ref{table:parameter:values}, 
and $u=0$, $u=20\%$, $u=40\%$, $u=60\%$, we have the following values 
for the basic reproduction number $\tilde{R}_0$ of Section~\ref{section:vaccine:constant}: 
$\tilde{R}_0=0.0647$, $\tilde{R}_0=0.0554$, $\tilde{R}_0=0.0484$, and $\tilde{R}_0=0.043$, 
respectively. From Theorem~\ref{theo3}, the disease free equilibrium $\Sigma_1$ 
of system \eqref{model:seir:normalized:vaccine:1} is locally asymptotically stable 
for the time delay $\tau=0$. From Theorem~\ref{theo4}, the endemic equilibrium point $\Sigma_V^+$ 
of system \eqref{model:seir:normalized:vaccine:1} is unstable for the time delay $\tau=0$.

In conclusion, there is an inverse proportional relationship between 
the fraction $u$ of susceptible individuals that are vaccinated and the 
number of exposed, infected, and recovered individuals: the greater the 
fraction of susceptible individuals that are vaccinated, the smaller 
the number of exposed, infected, and recovered individuals would be, 
and vice versa (see Figure~\ref{fig1:tau:fixe}). Moreover, there is a 
directly proportional relationship between the transfer time delay $\tau$ 
from the class of susceptible individuals to the class of infected individuals 
and the number of exposed, infected, and recovered individuals: 
the greater the time delay, the greater the number of exposed, 
infected, and recovered individuals would be, and vice versa 
(see Figure~\ref{fig1:u:fixe}).

\vspace{-6pt} 


\authorcontributions{Conceptualization, M.A.Z., C.J.S. and D.F.M.T.; 
methodology, M.A.Z., C.J.S. and D.F.M.T.; 
software, M.A.Z.; 
validation, M.A.Z., C.J.S. and D.F.M.T.; 
formal analysis, M.A.Z., C.J.S. and D.F.M.T.; 
investigation, M.A.Z., C.J.S. and D.F.M.T.; 
writing---original draft preparation, M.A.Z., C.J.S. and D.F.M.T.; 
writing---review and editing, M.A.Z., C.J.S. and D.F.M.T.; 
visualization, M.A.Z.; 
supervision, C.J.S. and D.F.M.T.
All authors have read and agreed to the published version of the~manuscript.}

\funding{This research was funded by FCT (Funda\c{c}\~ao para a Ci\^encia e a Tecnologia), 
grant number UIDB/04106/2020 (CIDMA). C.J.S. was also supported by FCT via the FCT 
Researcher Program CEEC Individual 2018 with reference CEECIND/00564/2018.}

\institutionalreview{Not applicable.}

\informedconsent{Not applicable.}

\dataavailability{Publicly available datasets were analyzed in this study. These data can be found here:
\url{https://raw.githubusercontent.com/pcm-dpc/COVID-19/master/dati-regioni/dpc-covid19-ita-regioni.csv}
(accessed on 14 August 2021).} 

\acknowledgments{The authors are very grateful to four reviewers 
for several constructive comments, suggestions and questions 
that helped them to improve their manuscript.}

\conflictsofinterest{The authors declare no conflict of interest. 
The funders had no role in the design of the study; in the collection, 
analyses, or interpretation of data; in the writing of the manuscript; 
or in the decision to publish the~results.} 


\end{paracol}

\reftitle{References}



\end{document}